\documentclass[floatfix,aps,prl,showpacs,amsmath,nofootinbib,
preprintnumbers,twocolumn]{revtex4}

\usepackage{graphicx}

\def\laq{~\raise 0.4ex\hbox{$<$}\kern -0.8em\lower 0.62
ex\hbox{$\sim$}~}
\def\gaq{~\raise 0.4ex\hbox{$>$}\kern -0.7em\lower 0.62
ex\hbox{$\sim$}~}

\def\beq{\begin{equation}}
\def\eeq{\end{equation}}
\def\bea{\begin{eqnarray}}
\def\eea{\end{eqnarray}}
\def\bean{\begin{eqnarray*}}
\def\eean{\end{eqnarray*}}

\def \ti {\widetilde}

\def \La {\Lambda}

\def \ga {\gamma}

\def \Om {\Omega}

   \def\be{\begin{equation}}
   \def\ee{\end{equation}}
   \def\ba{\begin{eqnarray}}
   \def\ea{\end{eqnarray}}

\def\laq{~\raise 0.4ex\hbox{$<$}\kern -0.8em\lower 0.62ex\hbox{$\sim$}~}
\def\gaq{~\raise 0.4ex\hbox{$>$}\kern -0.7em\lower 0.62ex\hbox{$\sim$}~}

\def\beq{\begin{equation}}
\def\eeq{\end{equation}}
\def\bea{\begin{eqnarray}}
\def\eea{\end{eqnarray}}

\def \ti {\widetilde}

\def \La {\Lambda}

\def \ga {\gamma}

\def \Om {\Omega}

    \def\be{\begin{equation}}
    \def\ee{\end{equation}}
    \def\ba{\begin{eqnarray}}
    \def\ea{\end{eqnarray}}

\newcommand{\eq}{\begin{equation}}
\newcommand{\eqx}{\end{equation}}
\newcommand{\eqn}{\begin{eqnarray}}
\newcommand{\eqnx}{\end{eqnarray}}

\newcommand{\Ups}{\Upsilon}

\begin{document}

\preprint{BA-TH 656-12}
\preprint{CERN-PH-TH/2012-188}
\preprint{LPTENS-12/24}

\title{Do stochastic inhomogeneities affect dark-energy precision measurements?}

\author{I. Ben-Dayan$^{1,2}$, M. Gasperini$^{3,4}$, G. Marozzi$^{5}$, F. Nugier$^{6}$ and G. Veneziano$^{5,7}$}

\affiliation{$^1$Canadian Institute for Theoretical Astrophysics, 60 St
George, Toronto ON, M5S 3H8\\
$^{2}$ Perimeter Institute for Theoretical Physics,
Waterloo, Ontario N2L 2Y5, Canada\\
$^{3}$Dipartimento di Fisica, Universit\`{a} di Bari, Via G. Amendola
173, 70126 Bari, Italy\\
$^{4}$Istituto Nazionale di Fisica Nucleare, Sezione di Bari, Bari, Italy\\
 $^{5}$ Coll\`ege de France, 11 Place M. Berthelot, 75005 Paris, 
 France\\
$^{6}$ Laboratoire de Physique Th\'eorique de l'\'Ecole Normale Sup\'erieure, CNRS UMR 8549, 24 Rue Lhomond, 75005 Paris, France\\
$^{7}$CERN, Theory Unit, Physics Department,  CH-1211 Geneva 23, Switzerland}


\begin{abstract}
The effect of  a stochastic background of cosmological perturbations on the luminosity-redshift relation is computed to second order through a recently proposed covariant and gauge-invariant light-cone averaging procedure. The resulting expressions are  free from both ultraviolet and infrared divergences, implying that such perturbations cannot mimic a sizable fraction of dark energy. Different averages are estimated and  depend  on the particular function of the luminosity distance being averaged. The energy flux,  being minimally affected by perturbations at large $z$, is proposed as the best choice for precision estimates of  dark-energy parameters. Nonetheless, its irreducible (stochastic) variance induces statistical errors on  $\Omega_{\Lambda}(z)$ typically lying in the few-percent range. 

\end{abstract}

\vspace {1cm}~


\maketitle

Establishing the existence of dark energy and determining its parameters is one of the central issues in modern cosmology.
Evidence for a sizable dark-energy component in the cosmic fluid comes from different sources: CMB anisotropies, models of large-scale-structure formation and, most directly, the luminosity redshift relation of Type Ia supernovae, used as standard candles.

In this latter case, on which we concentrate our attention, the analysis is usually made in the simplified context of a homogeneous and isotropic (FLRW) cosmology. The issue has then been raised about whether inhomogeneities may affect the conclusion of such a naive analysis.
Inhomogeneous models in which we occupy a privileged position in the Universe, for instance, can mimic dark energy (as first pointed out in \cite{LTB}), but look both unrealistic and highly fine-tuned. More interestingly, we should address this question in the presence of stochastically isotropic and homogeneous perturbations of the kind predicted by inflation. 
We present here the main ideas and results of such a study, 
while its detailed derivation and discussion is presented in \cite{BGMNV3} 
and in a forthcoming paper \cite{BGMNV4}.

There is by now  general agreement that super-horizon perturbations cannot mimic dark-energy effects \cite{super}. By contrast,
 the impact of sub-horizon perturbations  is still unsettled \cite{sub,Kolb,CU} owing to the appearance of ultraviolet divergences~\footnote{See \cite{MU} for the possible observational impact  of such ultraviolet divergences  on the anisotropy of the Hubble flow.}
 while computing their ``backreaction"  on certain classes of large-scale averages  \cite{Kolb,CU}. The possibility that these effects may simulate a substantial fraction of dark energy, or that they may at least play some role in the context of near-future precision cosmology, has to be seriously considered.

In order to address these issues we have studied the luminosity-redshift relation in a spatially-flat $\Lambda$CDM model, perturbed by a stochastic background of inhomogeneities. 
The  luminosity distance $d_L$ now depends on the redshift $z$ as well as on the angular coordinates of the sources, and must be inserted in an appropriate light-cone and {\em ensemble} average \cite{GMNV,BGMNV1}. Unlike the analyses in  \cite{Kolb,CU}, we find a result always free from ultraviolet divergences and with no significant infrared contributions either.
As a consequence, corrections are typically small, certainly too small to mimic a sizeable fraction of dark energy. However, interestingly enough, both their size and their $z$-dependence strongly depend on the particular function of $d_L$ being averaged. 

We find, in particular, that the energy flux $\Phi \sim d_L^{-2}$  is practically unaffected by inhomogeneities, 
 while the most commonly used variables (like the distance modulus $\mu \sim 5 \log_{10} d_L$) may receive much larger corrections. This creates (at least in principle) intrinsic ambiguities in the measure of the dark-energy parameters, unless the backreaction of stochastic inhomogeneities is properly taken into account.  
 Actually,  the advantages of flux averaging for minimizing biases on dark-energy parameters was first pointed out in \cite{P1}, where it was shown how the binning of data in appropriate redshift intervals can reduce the bias due to systematic effects such as weak lensing. It is intriguing that the preferred role played by the flux variable also comes out in this paper where we perform a completely different averaging procedure, {\em at fixed redshift}. Our conclusions are not due to a binning of data, but to an application of our covariant space-time average to different functions of the luminosity distance.

Let us start by recalling the standard expression for the luminosity distance in an unperturbed  flat $\Lambda$CDM model, with present fractions of critical density $\Om_\La$  and $\Om_m=1-\Om_\La$:
\beq
d_L^{FLRW}(z)=  {1+z\over H_0} \int_0^{z} {dz' \over\left[\Om_\La +\Om_{m} (1+z')^{3}\right]^{1/2}}.
\label{1}
\eeq
Consider now the  expression for $d_L$ in the corresponding perturbed geometry. Combining light-cone and ensemble averages (denoted, respectively, by brackets and over-bars),  we can write the averaged result in the form:
\beq
\overline{\langle d_L \rangle}(z)=d_L^{FLRW} \left[1+f_d(z)\right]\, ,
\label{2}
\eeq
where $f_d(z)$ represents the ``backreaction"  on $d_L$ due to  inhomogeneities. For consistency, $d_L$ has to be computed (at least) up to the second perturbative order since {\em ensemble} averages of first-order quantities are vanishing for stochastic perturbations. 
In particular, backreaction terms arise also from correlations between the inhomogeneities present  in the averaged variable and in the covariant integration measure. Therefore, a consistent perturbative calculation requires the  inclusion of linear second-order contributions, since they are  of the same order as the above quadratic first-order terms (see also \cite{BGMNV1}, Sect. 4). 
A detailed computation of $f_d(z)$ would thus enable to extract  the ``true" value of the dark-energy parameters from the  measurement of $\overline{\langle d_L \rangle}(z)$ after taking the correction into account. 

However, as already stressed in \cite{BGMNV1}, given  the covariant (light-cone) average of a perturbed (inhomogeneous) observable $S$  the average of a generic function of this observable  differs, in general, from the function of its average, i.e. $\overline{\langle F(S) \rangle} \not= F(\overline{\langle S\rangle})$. Expanding the observable  to second order as $S=S_0+S_1+S_2+ \cdots$, one finds:
\beq
\overline{\langle F(S) \rangle} = F(S_0)+ F'(S_0)
\overline{\langle S_1+S_2  \rangle}+F''(S_0) \overline{\langle S_1^2/2 \rangle}
\label{3}
\eeq
where $\overline{\langle S_1\rangle}\neq 0$ as a consequence of the ``induced backreaction" terms (see \cite{BGMNV1}, Sect. 4).
Thus different functions of the luminosity distance are differently affected by the inhomogeneities, and  require different ``subtraction" procedures.  Finding the function that minimizes the backreaction will help of course for a precision estimate of the cosmological parameters. One of the main claims of this paper is the  identification of such an optimal observable with the energy  flux $\Phi= L/(4 \pi d_L^2)$ received from a standard candle of luminosity $L$ located on the observer's past light-cone. We now illustrate how we have performed such a calculation.

The average value of $\Phi$, obviously controlled by the average of $d_L^{-2}$, has to be carried out on the past light-cone of the observer, at a fixed redshift $z$, using the  gauge-invariant prescription introduced in \cite{GMNV}. This is most conveniently done \cite{GMNV,BGMNV1} in the so-called geodesic light-cone gauge (GLC), where the metric depends on six arbitrary functions ($\Ups, U^a, \ga_{ab}$, $a,b=1,2$), 
and the line-element takes the form (with $\ti{\theta^1}= \ti{\theta}, \ti{\theta^2} = \ti{\phi}$):
\beq
\! ds^2\! = \!\Ups^2 dw^2\!-\!2\Ups dw d\tau\!+ \gamma_{ab}(d\ti{\theta}^a\!-U^a dw)(d\ti\theta^b\!-U^b dw).
\label{4}
 \eeq
The correspondence between the GLC gauge and the spatially flat FLRW geometry is \cite{GMNV}: $\tau=t$, $w=r+\eta$, $\Ups =a(t)$, $U^a=0$ and $\ga_{ab} d \tilde{\theta}^a d \tilde{\theta}^b = a^2 r^2 (d \tilde{\theta}^2+ \sin^2 \tilde{\theta} d \tilde{\phi}^2)$, 
where $\eta$ is the conformal-time coordinate ($d \eta= dt/a$). 

In the GLC gauge the past light-cone is defined by the condition $w = w_0=$ const, and the redshift is given by:
\be
1+z= 
{\Ups(w_0,\tau_0,\ti \theta^a)}/{\Ups(w_0,\tau,\ti \theta^a)}.
\label{5}
\ee
Furthermore, the luminosity distance of the source is simply expressed as \cite{BGMNV1} $d_L= (1+z)^2 \ga^{1/4} (\sin \ti \theta)^{-1/2}$,
yielding the following exact result \cite{BGMNV3}:
\beq
\langle d_L^{-2} \rangle(z, w_0) = {{4 \pi}(1+z)^{-4}\over \int {d^2 \ti{\theta}^a}\sqrt{ \gamma (w_0, \tau(z,\ti{\theta^a}),\ti{\theta}^b)}},
\label{6}
 \eeq
where $\ga= \det \ga_{ab}$, and $\tau(z,\tilde{\theta}^a)$ is obtained by solving Eq. (\ref{5}). The above expression has a simple physical interpretation: the averaged flux, for a given $z$,  is inversely proportional to the proper area (computed with respect to the metric ({\ref{4})) of the surface lying on our past light-cone at the given value of $z$. Flux conservation is probably at the basis of the particular simplicity of this average and of its minimal deviation from the homogeneous value.

To compute this quantity in the perturbed geometry of our interest, we need  to express it in a gauge where the stochastic background of cosmological perturbations is explicitly known up to second order.  To this purpose, we can use the standard Poisson gauge where we include first and second-order scalar perturbations,  neglecting their  tensor and vector counterparts (see  \cite{BGMNV3} for a discussion of this point). 
 Performing  the relevant transformations to second order we arrive at the following analogue of (\ref{2}):
\beq
\!\!\!\overline{\langle d_L^{-2} \rangle}\!=\!(d_L^{FLRW})^{-2}\overline{(I_\Phi(z))^{-1}}
\!\equiv\!
(d_L^{FLRW})^{-2} \left[1\!+\!f_\Phi(z) \right],
\label{7} 
\eeq
where $I_\Phi$ has in general the following structure:
\beq
I_\Phi(z) =
\int {d \tilde{\phi} d\tilde{\theta} \sin \tilde{\theta} \over 4 \pi}  
\Big[1+{\cal I}_1+{\cal I}_{1,1}+{\cal I}_2 \Big] (\tilde{\theta}, \tilde{\phi}, z).
\label{8}
\eeq
Here ${\cal I}_1,~ {\cal I}_{1,1},~ {\cal I}_2 $ are, respectively, the first-order, 
quadratic first-order, and
genuine second-order  contributions of our stochastic fluctuations. After solving the relevant perturbation equations \cite{BMR} they can all be expressed in terms of  the first-order Bardeen potential $\Psi(x, \eta)$. 
Using the stochastic properties of this perturbation, and expanding in Fourier modes $\Psi_k(\eta)$, we can then obtain 
an expression for $\overline{(I_\Phi)^{-1}}$ where  first-order contributions drop out because of the {\em ensemble} average, 
and the scalar perturbations only appear through the so-called dimensionless power spectrum, ${\cal P}(k, \eta)= (k^3/2 \pi^2)|\Psi_k(\eta)|^2$. 

Unfortunately, $\overline{(I_\Phi)^{-1}}$ contains integrals over null geodesics lying on the past light-cone of the given observer (see \cite{BGMNV1}, Sect. 3.2), which get intertwined with the  time-dependence of $\cal P$, forcing us to proceed with an approximate numerical integration. This will be done below, after inserting (as an instructive 
example) an illustration of the limiting CDM case, where all integrals but the one over $k$ can be done analytically thanks to the time-independence of ${\cal P}$~(\cite{BGMNV1}, Sect. 5). 

\begin{figure}[t!]
\centering
\includegraphics[width=8cm]{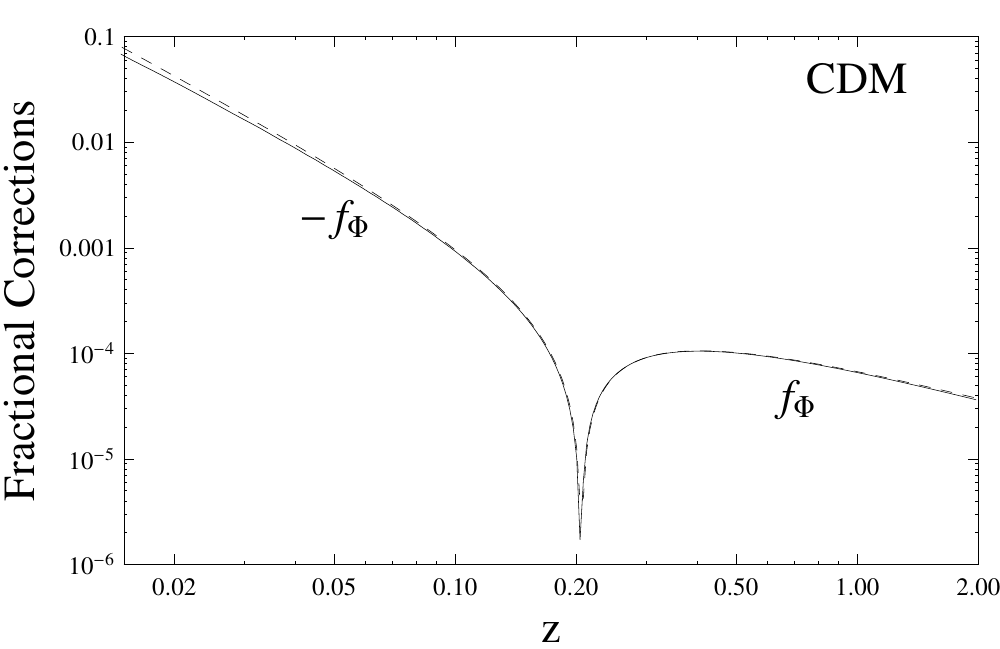}
\centering
\caption{The fractional correction $f_\Phi$ of Eq. (\ref{9}) (solid curve), compared with the same quantity given to leading order by Eq. (\ref{10}) (dashed curve), in the context of an inhomogeneous CDM model. We have used for ${\cal P}(k)$ the inflationary scalar spectrum with the WMAP parameters \cite{WMAP7} and  the transfer function  given in \cite{EH} (see also \cite{BGMNV1}).  The plotted curve refers,
as an illustrative example, to an UV cutoff $k_{UV}=1 {\rm Mpc}^{-1}$. }
 \label{f1}
\end{figure}

In that case the result can be written in the form
\beq
f_\Phi(z) = 
  \int_0^{\infty} \frac{d k}{k} \, {\cal P}(k) \Big[f_{1,1}(k,z) +  f_{2}(k,z) \Big] , 
\label{9}
\eeq
where $f_{1,1}$ and  $f_{2}$ are complicated --but known-- analytic functions of their arguments \cite{BGMNV4}. Furthermore, the leading contribution in the region of $z$ relevant for 
dark-energy phenomenology comes from terms of the type $f(k, z) \sim (k/ {\cal H}_0)^2 \ti f(z)$, where $ {\cal H}_0$ is the present Hubble scale. We can then write, to a very good accuracy,
\beq
f_\Phi(z) \simeq \Big[\ti f_{1,1}(z) +  \ti f_{2}(z) \Big] \int_0^{\infty} \frac{d k}{k} \,\left(  \frac{k}{  {\cal H}_0}\right)^2{\cal P}(k)  , 
\label{10}
\eeq 
where an explicit calculation gives \cite{BGMNV4}:
\bea
\!\!\!\!\! \!\!\!\!\!
\ti f_{1,1}(z) &=&   \frac{10-12 \sqrt{1+z}+5 z \left(2+\sqrt{1+z}\right)}{27 (1+z) \left(\sqrt{1+z} -1 \right)^2},  
\label{11}\\
\!\!\!\!\! \!\!\!\!\!
\ti f_2(z) &=&- \frac{1}{189} \left[\frac{2-2 \sqrt{1+z}+z \left(9-2 \sqrt{1+z}\right)}{(1+z) \left(\sqrt{1+z} - 1\right)}\right]\!.
\label{12}
\eea
The absolute value (and sign) of $f_\Phi(z)$ are illustrated in Fig. \ref{f1}, showing the accuracy of the leading order terms (\ref{10}) and confirming that the backreaction of a realistic  spectrum of stochastic perturbations induces  negligible corrections to the averaged flux at large $z$ (the larger corrections at small $z$, due to ``Doppler terms", has been discussed in \cite{BGMNV1}). 
In addition, it shows that, in any case, such corrections  have the wrong $z$-dependence (in particular change sign at some $z$) to simulate even a tiny  dark-energy component. 
 For the  considered spectrum (behaving as $k^{n_s-5} \log^2 k$ at large $k$, see \cite{EH}) the spectral integral is convergent  and very weakly  sensitive to the chosen value of the UV cutoff \cite{BGMNV1} representing here the limit of validity of our
perturbative approach.

\begin{figure}[t]
\centering
\includegraphics[width=8cm]{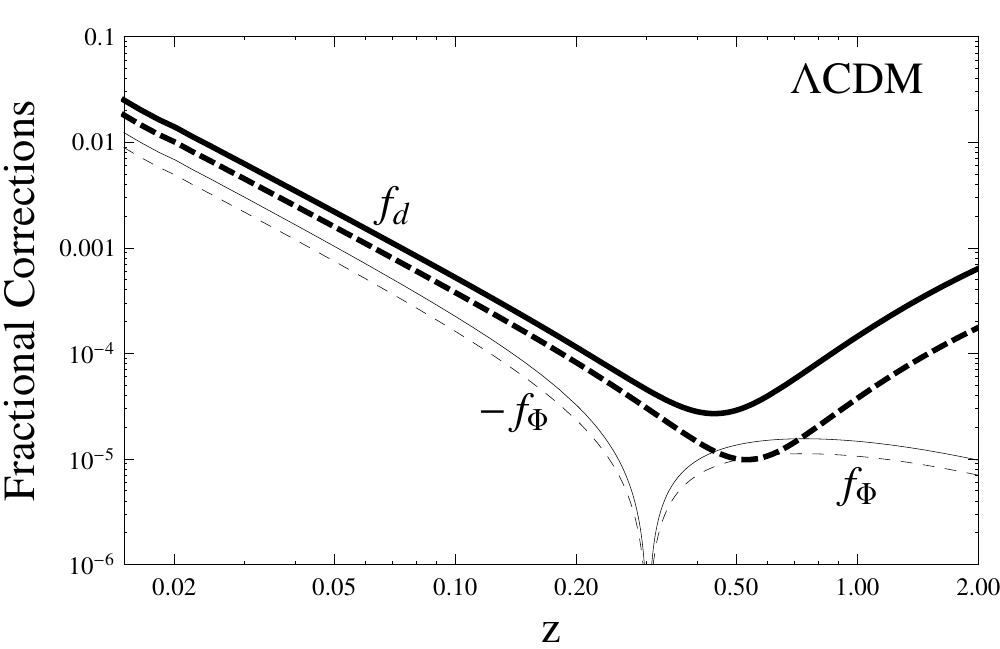}
\centering
\caption{The fractional correction to the flux $f_\Phi$ of Eq. (\ref{7}) (thin curves) is compared with the fractional correction to the luminosity distance $f_d$  of Eq. (\ref{13}) (thick curves), for a $\La$CDM model with $\Om_\La=0.73$. We have used two different cutoff values: $k_{UV}=0.1 {\rm Mpc}^{-1}$ (dashed curves) and $k_{UV}=1 {\rm Mpc}^{-1}$ (solid  curves);  the  spectrum is the same as that of Fig. \ref{f1}, adapted to $\La$CDM.}
\label{f2}
\end{figure}

We now come  to the more realistic $\La$CDM case, where the $f_\Phi$ correction should be obtained by a full numerical integration of Eqs. (\ref{7},\ref{8}). For simplicity, we will only take into account those terms giving the leading ($k^2$-enhanced)  contributions in the CDM case. For $\La$CDM we can generally expect a smaller correction due to the fact that the spectrum is now  suppressed, at large $k$, by a lower value of the equality scale $k_{\rm eq}$ \cite{EH}. This is confirmed by the explicit numerical result for $|f_\Phi|$ presented in Fig. \ref{f2}. 
The small value of $|f_\Phi|$ at large $z$  leads us to conclude that the averaged flux is a particularly appropriate quantity for extracting from the observational data the ``true" cosmological parameters. As we are going to see now, the situation is somewhat different for other functions of $d_L$. 

Indeed, let's apply the general result (\ref{3}) to the flux variable, $S= \Phi$, and consider two important examples: $F(\Phi)= \Phi^{-1/2} \sim d_L$, and $F(\Phi)= -2.5 \log_{10}\Phi + {\rm const} \sim \mu$ (the distance modulus). For the luminosity distance, following the notations of Eq. (\ref{2}) and using the general result (\ref{3}), we obtain:
\beq
f_d = -(1/2)f_\Phi+(3/8)\overline{\langle \left(\Phi_1/\Phi_0\right)^2 \rangle}.
\label{13}
\eeq
Similarly, for the distance modulus we obtain:
\beq
\overline{\langle \mu \rangle}- \mu^{FLRW}= - 1.25 (\log_{10}e)
\Big[ 2f_\Phi-\overline{\langle \left(\Phi_1/\Phi_0\right)^2 \rangle}
\Big],
\label{14}
\eeq
where $f_\Phi$ is defined in Eq. (\ref{7}). 

As clearly shown by the two above equations, the corrections to the averaged values of $d_L$ and $\mu$ are qualitatively different from those of the flux (represented by $f_\Phi$), because of the extra  contribution (inevitable for any non-linear function of the flux) proportional to the square of the first-order fluctuations. As mentioned before, the averaged flux corrections have leading spectral contributions of the type $k^2 {\cal P}(k)$; on the contrary, the new corrections to $d_L$ and $\mu$ are due to the so-called ``lensing effect", they dominate at large $z$, and have leading spectral contributions of the type $k^3 {\cal P}(k)$ (as already discussed in \cite{BGMNV1}). The explicit numerical integration, reported in Fig. 2,   confirms that, as a result, $|f_\Phi| \ll f_d$ at large $z$. We stress that even the $k^3$-enhanced contributions are UV-finite for the case under consideration.

We also stress that our results concerning the effects of lensing are in good agreement with previous estimates of the bias on supernova observables \cite{P2} and other cosmological parameters \cite{P3} induced by weak-lensing magnification effects. Unlike in those papers, however, our general approach automatically includes (and estimates the effects of) all possible corrections due to the stochastic fluctuations of the cosmological background, to second order, for all given functions of the flux (or of $d_L$). In fact, as discussed in detail in \cite{BGMNV3, BGMNV4}, the fractional correction $f_d$ includes, besides the lensing effect, also Doppler, Sachs-Wolfe,  
integrated Sachs-Wolfe, frame-dragging effects, etc. 

Let us now briefly discuss to what extent the enhanced corrections due to the squared first-order fluctuations can affect the determination of the dark-energy parameters if quantities other than the flux are used in the fits. To this purpose we consider the much used (average of the) distance modulus given in Eq. (\ref{14}), referred as usual to the homogeneous Milne model with $\mu^M= 5 \log_{10} [(2+z)z/(2H_0)]$.  In  Fig. \ref{f3} we compare  the averaged value $ \overline{\langle \mu \rangle} -\mu^M$ with the corresponding expression in a homogeneous $\La$CDM model with different values of $\Om_\La$. We also show the expected dispersion around the averaged result, represented by the square root of the variance \cite{BGMNV1}. The latter is given  by:
\beq
\sqrt{ \overline{\langle \mu^2 \rangle}- \left( \overline{\langle \mu \rangle}\right)^2}= \pm 2.5 (\log_{10} e) \sqrt{\overline{\langle \left(\Phi_1/\Phi_0\right)^2 \rangle}};
\label{15}
\eeq
while for the flux we simply find:
\beq
\sqrt{ \overline{\langle \left(\Phi/\Phi_0\right)^2 \rangle}- \left( \overline{\langle \Phi/\Phi_0 \rangle}\right)^2}= \pm \sqrt{\overline{\langle \left(\Phi_1/\Phi_0\right)^2 \rangle}}\, .
\label{16}
\eeq

 As illustrated in Fig. 3, we find that, even for the distance modulus, the effect of inhomogeneities on the average only affects the determination of 
$\Om_\La$ at the third decimal figure (see also Fig. 2), at least for the inflationary power spectrum with the $\Lambda$CDM transfer function of \cite{EH}: in that case, the curves for $\overline{\langle \mu \rangle}$ and $\mu^{\rm FLRW}$ are practically coincident at large $z$. We have considered other spectra which take into account non-linear effects and have more power at short scales, like those  obtained following \cite{Smith}. Using such spectra only  affects very mildly the $k^2$-enhanced terms (hence the flux) while they  increase the corrections wherever the $k^3$-enhanced lensing terms play a major role. In particular, the  variance due to the fluctuations, which is already at the few-$\%$ level at large $z$ for the power spectrum of \cite{EH}  (see Fig. \ref{f3}), can be further increased \cite{BGMNV4}. Note  that, even for these improved spectra, all our integrals are still free of UV divergences since, in any case, $\cal{P}$ falls faster than $k^{-3}$ (i.e. the matter density constrast spectrum grows slower than $k$).

\begin{figure}[t]
\centering
\includegraphics[width=8cm]{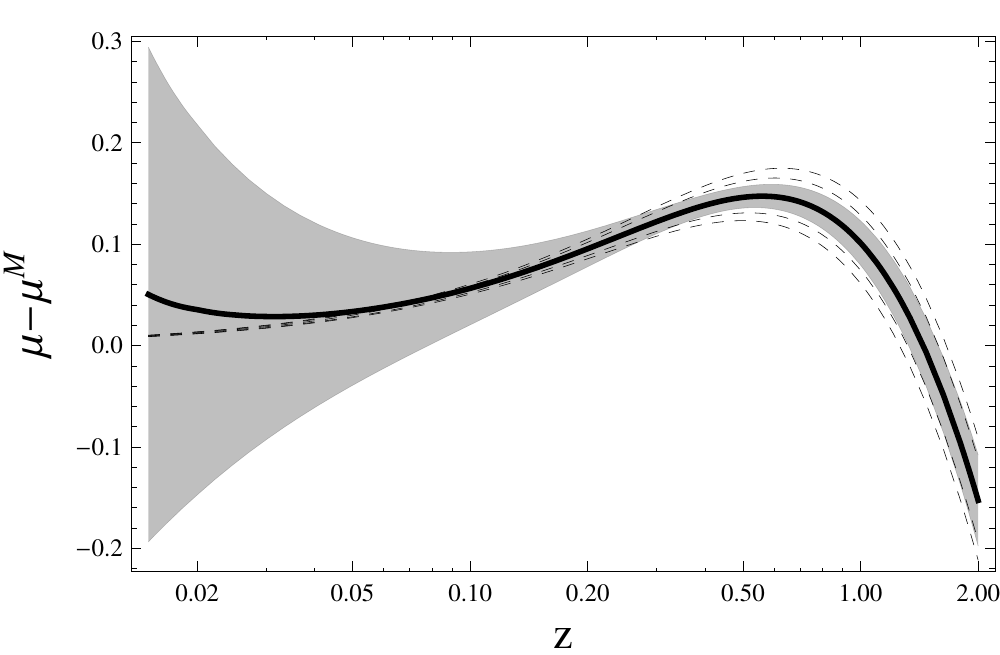}
\centering
\caption{The averaged  distance modulus 
$ \overline{\langle \mu \rangle} -\mu^M$ (thick solid curve), and its dispersion of Eq. (\ref{15}) (shaded region) are computed for $\Om_\La=0.73$ and compared with the homogeneous value  for the unperturbed $\La$CDM models with  
$ \Om_\La= 0.69$, $0.71$, $0.73$, $0.75$, $0.77$ (dashed curves). We have used  $k_{UV}=1\rm{Mpc}^{-1}$ and the same spectrum as in Fig. \ref{f2}.}
\label{f3}
\end{figure}

Our main conclusions can be summarized as follows:

(1) Dealing directly with the experimentally measured luminosity-redshift relation within a gauge-independent approach leads to results for the fractional corrections to the averaged variables and the corresponding variances  
 which are automatically free from UV (and IR) divergences for {\it any} function of the luminosity distance. 
 This can be contrasted with the case of more formal space-like averages \cite{Kolb,CU} for which the physical interpretation of the results may have no direct relation with the observed cosmic acceleration (first reference in \cite{sub}) and, as shown in \cite{CU}, the accidental cancellation of UV divergences is strongly dependent on the observable considered.
 
(2) The actual value of the backreaction  strongly depends on the quantity being averaged.  It turns out to be minimal for the flux $\Phi$, which is also practically insensitive to the short-distance behaviour of the power spectrum. Therefore, the flux stands out as the safest observable for precision cosmology.  For other observables, such as the distance modulus,  the backreaction is considerably  larger  and is more sensitive to the spectrum used.

(3) The dispersion due to stochastic fluctuations is much larger than the backreaction itself, implying an irreducible scatter of the data that may limit to the percent level (see Fig. \ref{f3}) the precision attainable on cosmological parameters because of the present limited statistics.

(4) We calculated here the full second order effect of stochastic perturbations and concluded that they cannot simulate a substantial fraction of dark energy. 
Possible contributions coming from the non-perturbative regime  on length scales much smaller than $1$ Mpc have still to be taken in consideration before final conclusions can be drawn. 

We wish to thank Ruth Durrer, Valerio Marra, Slava Mukhanov, Misao Sasaki  and Roman Scoccimarro for stimulating discussions.
The research of IBD at Perimeter Institute is supported by the
Government of Canada through Industry Canada and by the Province of Ontario through the Ministry of Research \& Innovation.

\end{document}